\documentclass[11pt]{article}
\usepackage[]{acl}
\usepackage[utf8]{inputenc}
\usepackage[T1]{fontenc}
\usepackage{xcolor}
\definecolor{darkblue}{rgb}{0, 0, 0.5}
\usepackage[colorlinks=true,citecolor=darkblue,linkcolor=darkblue,urlcolor=darkblue]{hyperref}
\usepackage{url}
\usepackage{booktabs}
\usepackage{amsfonts}
\usepackage{amsmath}
\usepackage{nicefrac}
\usepackage{microtype}
\usepackage{xcolor}
\usepackage{graphicx}
\usepackage{float}
\usepackage{enumitem}
\usepackage{multirow}

\title{SemanticAudio: Audio Generation and Editing in Semantic Space}

\author{
  Zheqi Dai$^{1}$,
  Guangyan Zhang$^2$,
  Haolin He$^1$,
  Xiquan Li$^3$,
  Jingyu Li$^2$,
  Chunyat Wu$^1$, \\
  Yiwen Guo$^{4,\star}$,
  Qiuqiang Kong$^{1,\star}$ \\
  $^1$The Chinese University of Hong Kong \quad
  $^2$LIGHTSPEED \\
  $^3$Shanghai Jiao Tong University \quad
  $^4$Independent Researcher \\
}

\begin{document}
\maketitle
\begin{abstract}
In recent years, Text-to-Audio Generation has achieved remarkable progress, offering sound creators powerful tools to transform textual inspirations into vivid audio. However, existing models predominantly operate directly in the acoustic latent space of a Variational Autoencoder (VAE), often leading to suboptimal alignment between generated audio and textual descriptions. In this paper, we introduce \textbf{SemanticAudio}, a novel framework that conducts both audio generation and editing directly in a high-level semantic space. We define this semantic space as a compact representation capturing the global identity and temporal sequence of sound events, distinct from fine-grained acoustic details. SemanticAudio employs a two-stage Flow Matching architecture: the \textbf{Semantic Planner} first generates these compact semantic features to sketch the global semantic layout, and the \textbf{Acoustic Synthesizer} subsequently produces high-fidelity acoustic latents conditioned on this semantic plan. Leveraging this decoupled design, we further introduce a training-free text-guided editing mechanism that enables precise attribute-level modifications on general audio without retraining. Specifically, this is achieved by steering the semantic generation trajectory via the difference of velocity fields derived from source and target text prompts. Extensive experiments demonstrate that SemanticAudio achieves competitive generation quality while providing significantly improved semantic alignment and enabling effective training-free audio editing. Demo available at: \href{https://semanticaudio1.github.io/}{\url{https://semanticaudio1.github.io/}}
\end{abstract}

\section{Introduction}

Text-to-Audio (TTA) Generation~\citep{liu2023audioldm,evans2024stableaudioopen,audioldm2-2024taslp} aims to synthesize diverse and high-fidelity auditory content directly from natural language textual prompts. This technology serves as a pivotal creative tool for applications including virtual reality, gaming, film post-production, and human-computer interaction. Recent years have witnessed a paradigm shift in this field, fueled by the scaling of data and model parameters alongside architectural innovations. In particular, the adoption of continuous generative objectives, exemplified by Diffusion Models and Flow Matching, has elevated the fidelity and controllability of synthesized audio.

Most mainstream TTA models perform modeling directly in the acoustic latent space, typically utilizing compressed representations from a Variational Autoencoder (VAE)~\citep{liu2023audioldm,audioldm2-2024taslp}. While this design excels at preserving low-level acoustic fidelity, it often falls short in high-level semantic understanding. These models frequently struggle to precisely capture the intent in textual prompts, resulting in insufficient \textit{alignment}—defined here as the accurate correspondence between the presence and sequence of auditory events and the text description.

Addressing this limitation requires a clear distinction between the \textit{semantic} and \textit{acoustic} levels of audio. In this work, we define semantics as the high-level conceptual content—specifically the identity, occurrence, and temporal sequence of sound events—as distinct from fine-grained acoustic details.
Audio signals exhibit significant semantic redundancy: high-level semantics are relatively compact and abstract compared to dense acoustic details. Drawing inspiration from two-stage semantic planning approaches in video generation, we hypothesize that directly modeling dense low-level representations in a high-dimensional acoustic latent space is suboptimal for achieving semantic alignment. Instead, the generation process should be decomposed: first accomplishing global content planning in a compact high-level semantic space, followed by the progressive synthesis of acoustic details.

Motivated by this insight, we propose \textbf{SemanticAudio}, a novel two-stage Flow Matching-based framework. The core innovation lies in performing the audio generation process via a high-level semantic space. First, a \textbf{Semantic Planner} generates compact semantic features from text to sketch the global event layout. Second, conditioned on these features, an \textbf{Acoustic Synthesizer} produces high-fidelity VAE latent representations. This design effectively addresses the limitations in high-level semantic modeling inherent in conventional acoustic-space approaches.

Beyond generation, we demonstrate that this decoupled architecture naturally extends to audio editing tasks. While attempting training-free text-guided editing~\citep{xu2023inversionfreeimageeditingnatural,kulikov2025floweditinversionfreetextbasedediting} with standard acoustic-space models, we observed unsatisfactory results due to the substantial semantic gap between text and acoustic latents. Leveraging SemanticAudio, we introduce a training-free editing mechanism that operates directly in the semantic space. By steering the generation trajectory via the difference of velocity fields derived from source and target prompts, we achieve precise attribute-level modifications. This stands in contrast to traditional audio editing methods~\citep{wang2023audit,liang2025audiomorphixtrainingfreeaudioediting}, which are typically limited to predefined operations such as addition or deletion. Our mechanism, by fully capitalizing on the advantages of semantic space, enables flexible, text-driven manipulation of high-level semantics on general audio without additional training.

The main contributions of this work are summarized as follows:\\
\noindent \textbf{SemanticAudio Framework}: We propose a two-stage framework comprising a \textbf{Semantic Planner} and an \textbf{Acoustic Synthesizer}. This architecture performs audio generation directly in a high-level semantic space, effectively decoupling content planning from acoustic synthesis.

\noindent \textbf{Superior Semantic Consistency}: By first \textbf{sketching the global event layout} in the semantic space, our method achieves substantial outperformance over existing mainstream methods in high-level semantic alignment between generated audio and textual prompts.

\noindent \textbf{Training-free Audio Editing}: We introduce a training-free mechanism that enables \textbf{flexible, text-driven manipulation of high-level semantics on general audio}. By directly steering the semantic ODE trajectory, this approach achieves versatile attribute-level modifications without requiring additional training or complex inversion steps.

\section{Related Work}
\label{sec:related}

\textbf{Text-to-Audio Generation} Recent advances in TTA generation have been driven by the scaling of latent diffusion models and Flow Matching frameworks. The prevailing paradigm involves compressing audio into an acoustic latent space via a Variational Autoencoder (VAE) trained on mel-spectrograms, followed by modeling the noise-to-data distribution within this space. Prominent approaches include AudioLDM~\citep{liu2023audioldm}, Make-An-Audio~\citep{huang2023makeanaudiotexttoaudiogenerationpromptenhanced}, AudioGen~\citep{kreuk2023audiogentextuallyguidedaudio}, and Tango~\citep{majumder2024tango2aligningdiffusionbased}. More recently, Flow Matching-based models such as MeanAudio~\citep{li2025meanaudiofastfaithfultexttoaudio} and LAFMA~\citep{guan2024lafmalatentflowmatching} have demonstrated improved training stability and sampling efficiency. Despite achieving high acoustic fidelity, these models predominantly operate directly in the high-dimensional acoustic latent space. This design conflates fine-grained acoustic details with high-level event logic, often leading to suboptimal semantic alignment, particularly regarding the temporal sequence and structure of sound events described in complex textual prompts.

\textbf{Semantic Representations in Audio} To bridge the semantic gap, prior works have explored various high-level audio representations. Early efforts utilized discrete semantic tokens, as seen in AudioLM~\citep{borsos2023audiolmlanguagemodelingapproach}, or continuous embeddings from contrastive models like CLAP~\citep{wu2024largescalecontrastivelanguageaudiopretraining} and AudioMAE~\citep{huang2023maskedautoencoderslisten}. However, these representations have largely served as auxiliary conditioning signals rather than the primary generation target. Furthermore, global descriptors like CLAP aggregate information into a single vector, losing the temporal granularity required for detailed event planning. In contrast, the recent Perception Encoder series, specifically PE-A-Frame~\citep{vyas2025pushingfrontieraudiovisualperception}, provides frame-level semantic embeddings trained with fine-grained audiovisual objectives. By capturing precise temporal alignment between audio frames and textual descriptions, PE-A-Frame offers a temporally rich semantic space suitable for the decoupled planning strategy we propose in this work.

\textbf{Audio Editing} Audio editing approaches typically fall into training-based or training-free categories. Training-based methods, such as Audit~\citep{wang2023audit} and RFM-Editing~\citep{gao2025rfmeditingrectifiedflowmatching}, rely on supervised learning with paired data (e.g., original/edited pairs) to learn specific instruction-following capabilities. While precise, they suffer from high data annotation costs and limited generalization to unseen instructions. Conversely, training-free methods leverage the inherent priors of pre-trained generative models. These often follow an inversion-based paradigm—exemplified by AudioMorphix~\citep{liang2025audiomorphixtrainingfreeaudioediting}—where the input audio is inverted to a noise latent and resampled with modified text conditions. However, these approaches are susceptible to inversion errors and struggle to disentangle semantic content from acoustic texture. While inversion-free editing via vector field composition (e.g., FlowEdit~\citep{kulikov2025floweditinversionfreetextbasedediting}) has proven effective in the image domain, its application to audio, particularly within a high-level semantic space, remains underexplored.

\textbf{Inspirations from Video and Image Generation} The concept of decoupling semantic planning from low-level synthesis has gained traction in visual generation. In video generation, SemanticGen~\citep{bai2025semanticgenvideogenerationsemantic} demonstrated that generating global layouts in a compact semantic space prior to pixel-level refinement significantly improves coherence in long sequences. Similar "coarse-to-fine" paradigms have been applied to image generation (e.g., RCG~\citep{li2024returnunconditionalgenerationselfsupervised} and TokensGen~\citep{ouyang2025tokensgenharnessingcondensedtokens}). SemanticAudio adapts this insight to the auditory domain, being the first framework to perform audio generation and editing directly within a continuous, high-level semantic space, effectively decoupling content planning from acoustic rendering.

\begin{figure*}[t]
    \centering
    \includegraphics[width=0.95\linewidth]{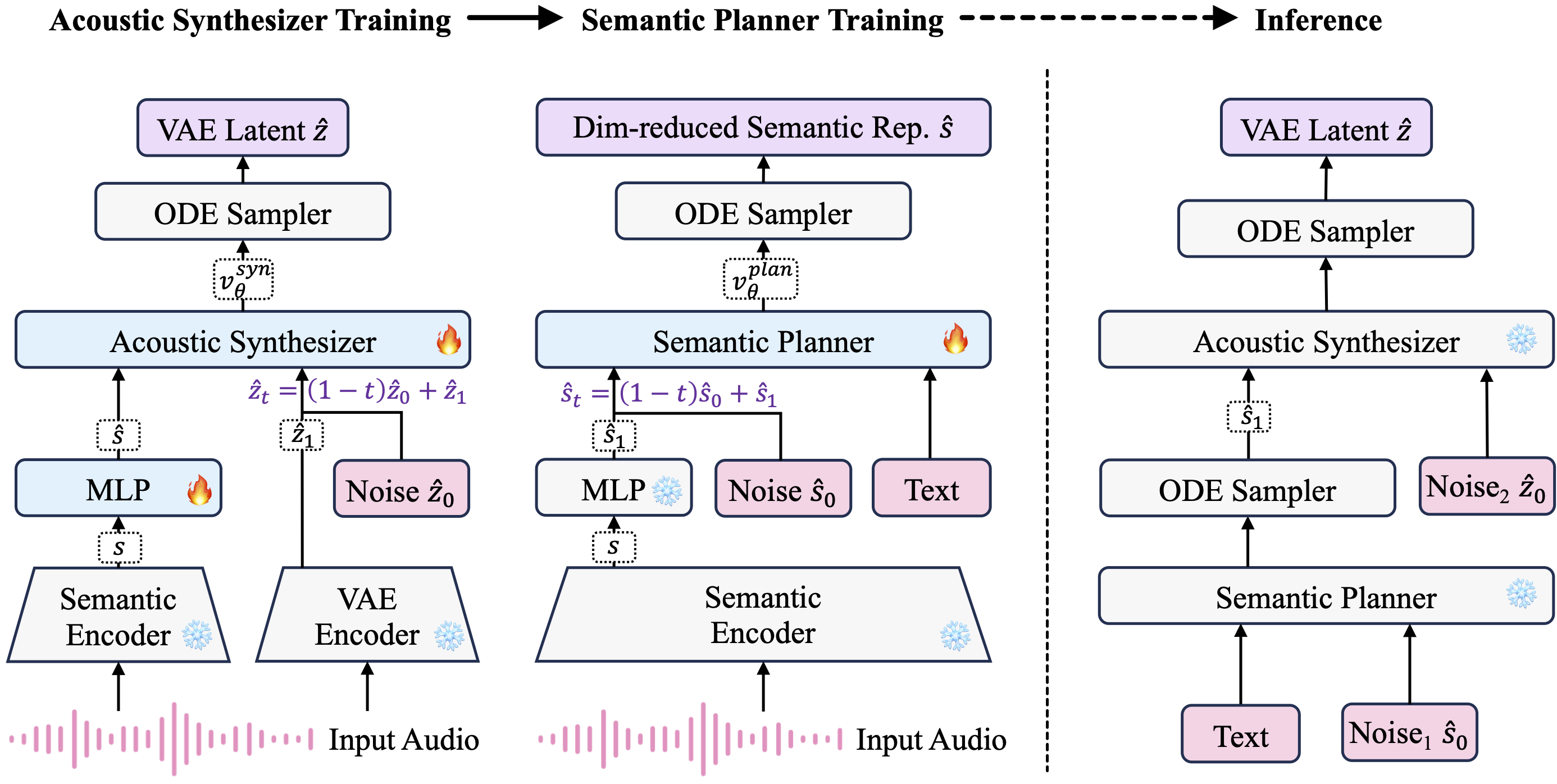}
    \caption{Overview of the SemanticAudio framework. The model employs a two-stage Flow Matching architecture: the \textbf{Semantic Planner} first generates low-dimensional semantic latents conditioned on text, followed by the \textbf{Acoustic Synthesizer} which produces high-fidelity acoustic latents for VAE decoding.}
    \label{fig:framework}
\end{figure*}

\section{SemanticAudio Framework}
\label{sec:framework}

In this section, we present the detailed architecture of the SemanticAudio framework. As illustrated in \autoref{fig:framework}, our framework effectively decouples text-to-audio generation into two distinct stages: (1) a \textbf{Semantic Planner} that sketches the global event layout in a compact semantic space, and (2) an \textbf{Acoustic Synthesizer} that produces high-fidelity acoustic details conditioned on the semantic plan. We first detail the representation spaces, followed by the design of the two generative stages.

\subsection{Pre-trained VAE and Semantic Representation}
\label{subsec:vae-semantic}

SemanticAudio builds upon a pre-trained variational autoencoder (VAE) and a semantic encoder to bridge raw audio waveforms and high-level semantics.

\textbf{Acoustic Representation.} SemanticAudio leverages a variational autoencoder (VAE) to compress a raw audio waveform into a compact acoustic latent space $z \in \mathbb{R}^{T \times C}$. Formally, the encoder $E_{\text{VAE}}$ maps the input waveform $a$ to a latent representation $z = E_{\text{VAE}}(a)$, where $T$ denotes the number of acoustic time steps and $C$ represents the channel dimension. The decoder $D_{\text{VAE}}$ reconstructs the audio from this latent, $\hat{a} = D_{\text{VAE}}(z)$, ensuring high perceptual fidelity. In this work, we adopt the pre-trained Descript Audio Codec (DAC)~\citep{kumar2023high} as our acoustic VAE.

\textbf{Semantic Representation.} To enable precise control over the temporal layout and content of sound events, we require a semantic encoder $E_{\text{sem}}$ capable of extracting continuous, frame-level embeddings $s \in \mathbb{R}^{N \times D}$. Here, $N$ corresponds to the number of semantic frames (determined by the frame rate of the encoder) and $D$ is the embedding dimension. Unlike global descriptors that aggregate information into a single vector (e.g., CLAP~\citep{elizalde2023clap}), frame-level representations are essential for preserving the fine-grained temporal structure required for event planning.

In this work, we adopt the pre-trained \textbf{Perception Encoder}~\citep{vyas2025pushingfrontieraudiovisualperception}. This model is trained via \textbf{fine-grained supervised contrastive learning} on large-scale audio-text datasets. By explicitly aligning audio frames with their corresponding textual descriptions, it excels at \textbf{capturing precise semantic-temporal correspondences}. This makes it uniquely capable of tasks requiring detailed event sequencing and distinguishing overlapping sound concepts, providing a robust foundation for our Semantic Planner.
To enable tractable modeling in the generative process, we introduce a lightweight MLP projection head $P_{\theta}$ that reduces these high-dimensional embeddings ($D=1024$) to a compact low-dimensional space:
\begin{equation}
\hat{s} = P_{\theta}(s) \in \mathbb{R}^{N \times d}, \quad d \ll D.
\end{equation}
The projection head $P_{\theta}$, which is randomly initialized, is trained jointly with the Acoustic Synthesizer and remains fixed during the subsequent training of the Semantic Planner. This design ensures that the reduced representations $\hat{s}$ preserve essential semantic content necessary for accurate acoustic synthesis.

\subsection{Semantic Planner: Text-to-Semantic Generation}
\label{subsec:planner}

The \textbf{Semantic Planner} is responsible for high-level content planning. It learns to generate low-dimensional semantic representations directly conditioned on text prompts, effectively sketching the global event layout.

Given a text prompt $y$, we extract complementary semantic conditions using two distinct \textbf{pre-trained} encoders. We employ the text encoder from CLAP~\citep{elizalde2023clap} to extract a global sentence embedding $c_g$, capturing the high-level atmosphere. Simultaneously, we use the Flan-T5~\citep{chung2022scalinginstructionfinetunedlanguagemodels} encoder to extract a sequence of token-level embeddings $c_d$, preserving fine-grained syntactic structures and dynamic instructions. To simplify notation, we denote the full textual conditioning set as $C = \{c_g, c_d\}$. These representations serve as dual inputs to ensure both global coherence and local precision.

The Semantic Planner is a Flow Matching model $\mathcal{F}_{\text{plan}}$ that learns a velocity field $v_{\theta}^{\text{plan}}(t, \hat{s}_t, C)$ to transport noise $\hat{s}_0 \sim \mathcal{N}(0, I)$ to the target semantic latent $\hat{s}_1$. The training objective follows the Flow Matching~\citep{lipman2023flow} loss:
\begin{equation}
\label{eq:fm_loss}
\mathcal{L}_{\text{FM}}^{\text{plan}} = \mathbb{E}_{t, \hat{s}_t, C} \left\| v_{\theta}^{\text{plan}}(t, \hat{s}_t, C) - (\hat{s}_1 - \hat{s}_0) \right\|^2,
\end{equation}
where $\hat{s}_t = (1-t)\hat{s}_0 + t\hat{s}_1$, and the target $\hat{s}_1 = P_{\theta}(E_{\text{sem}}(a_{\text{gt}}))$ is computed using the frozen projection head $P_{\theta}$ learned during Acoustic Synthesizer training. Here, $a_{\text{gt}}$ denotes the target ground-truth audio waveform from the training pair, and $E_{\text{sem}}$ is the fixed semantic encoder.

During inference, we sample $\hat{s}_0 \sim \mathcal{N}(0, I)$ and integrate the learned velocity field from $t{=}0$ to $t{=}1$ using the Euler method with $N$ discrete steps:
\begin{equation}
\label{eq:euler}
\hat{s}_{t+\Delta t} = \hat{s}_t + v_{\theta}^{\text{plan}}(t, \hat{s}_t, C) \cdot \Delta t, \quad \Delta t = \tfrac{1}{N},
\end{equation}
yielding the planned semantic features $\hat{s}_1$.

\subsection{Acoustic Synthesizer: Semantic-to-Acoustic Synthesis}
\label{subsec:synthesizer}

The \textbf{Acoustic Synthesizer} bridges abstract semantic plans and concrete auditory signals. Conditioned on the semantic features $\hat{s}_1$, it learns to synthesize high-fidelity acoustic latents $z_1 \in \mathbb{R}^{T \times C}$.

\textbf{Training Strategy.} A critical aspect of our framework is that \textbf{the Acoustic Synthesizer is trained prior to the Semantic Planner}. We jointly optimize the synthesizer and the projection head $P_{\theta}$. This ensures that the projected semantic features $\hat{s} = P_{\theta}(E_{\text{sem}}(a_{\text{gt}}))$ retain sufficient information for reconstruction while discarding redundant noise. Once trained, $P_{\theta}$ is frozen to provide target labels for the Semantic Planner.

\textbf{Modeling.} The synthesizer adopts the same Flow Matching formulation as the Semantic Planner (\autoref{eq:fm_loss}). It learns a velocity field $v_{\theta}^{\text{syn}}(t, z_t, \hat{s}_1)$ to map noise $z_0$ to the ground-truth acoustic latents $z_1 = E_{\text{VAE}}(a_{\text{gt}})$, conditioned on $\hat{s}_1$.

\textbf{Inference.} The full generation pipeline is executed sequentially: we first generate the semantic plan $\hat{s}_1$ using the Semantic Planner, which then serves as the condition for the Acoustic Synthesizer to generate $z_1$. Finally, the waveform is reconstructed via the VAE decoder $\hat{a} = D_{\text{VAE}}(z_1)$.

\begin{figure*}[t]
    \centering
    \includegraphics[width=0.95\linewidth]{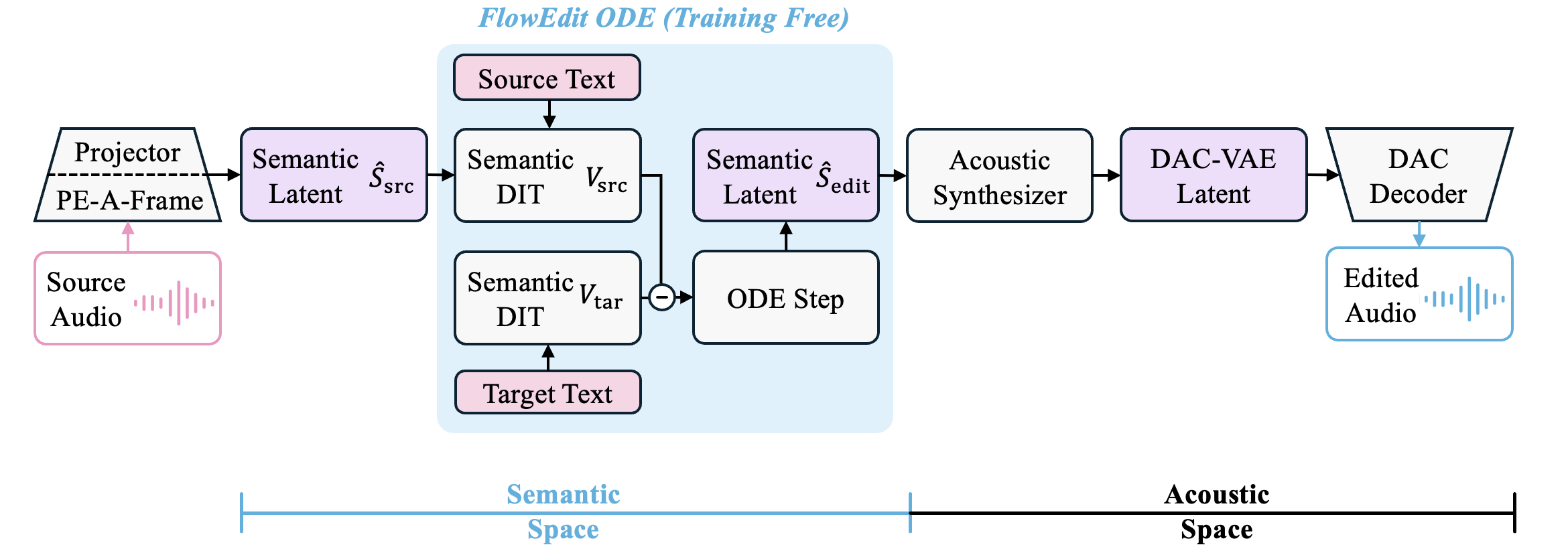}
    \caption{Overview of our training-free text-guided audio editing method. The process leverages the pre-trained velocity fields of the \textbf{Semantic Planner} to perform semantic-level editing in the low-dimensional latent space via difference velocity integration, followed by high-fidelity reconstruction using the \textbf{Acoustic Synthesizer}. The method requires no additional training, inversion, or optimization.}
    \label{fig:editing}
\end{figure*}

\subsection{Training-Free Text-Guided Audio Editing}
\label{subsec:editing}

A key advantage of our decoupled SemanticAudio framework is its inherent capability for \textbf{training-free audio editing}. Unlike pixel- or acoustic-space editing methods that often struggle to disentangle semantic content from background noise, our approach operates directly on the high-level semantic layout. This allows users to modify specific auditory events while preserving the underlying temporal structure, all without requiring model fine-tuning.

Building upon this insight, we introduce a mechanism inspired by FlowEdit~\citep{kulikov2025floweditinversionfreetextbasedediting}, as shown in \autoref{fig:editing}. It directly leverages the velocity fields learned by the \textbf{Semantic Planner} to perform precise semantic-level modifications, while the \textbf{Acoustic Synthesizer} ensures high-fidelity acoustic reconstruction.

Given a source audio $a_{\text{src}}$ and its semantic latent $\hat{s}_{\text{src}}$, we define the editing trajectory using a \textbf{Delta Velocity Field} $v_{\Delta}^t$. This field represents the directional difference between the Semantic Planner's velocity fields conditioned on the target ($C_{\text{tgt}}$) and source ($C_{\text{src}}$) prompts:
\begin{equation}
v_{\Delta}^t(\hat{s}^t, t) = v_{\theta}^{\text{plan}}(\hat{s}^t, t, C_{\text{tgt}}) - v_{\theta}^{\text{plan}}(\hat{s}^t, t, C_{\text{src}}).
\end{equation}
where $C_{\text{src}}$ can be the source text embedding or \textbf{null conditioning} if the source text is unavailable.

In practice, to ensure stability against stochastic variations, we approximate $v_{\Delta}^t$ by averaging over $N_{\text{avg}}$ noisy realizations at each timestep:
\begin{equation}
v_{\Delta}^t \approx \frac{1}{N_{\text{avg}}} \sum_{i=1}^{N_{\text{avg}}} \left[ v_{\theta}^{\text{plan}}(\hat{s}_{\text{tgt},i}^t, t, C_{\text{tgt}}) - v_{\theta}^{\text{plan}}(\hat{s}_{\text{src},i}^t, t, C_{\text{src}}) \right].
\end{equation}

Starting from the source semantic latent $\hat{s}^{1} = \hat{s}_{\text{src}}$, we integrate this delta field backward to $t=0$ using standard discrete steps (e.g., Euler method) to obtain the edited semantic latent $\hat{s}_{\text{edit}}$. Finally, $\hat{s}_{\text{edit}}$ is decoded by the \textbf{Acoustic Synthesizer} into the final audio.

\section{Experiments}
\label{sec:exp}
In this section, we empirically evaluate SemanticAudio on three primary tasks: text-to-audio generation, text-to-audio benchmark evaluation, and training-free semantic editing. We aim to verify our core hypothesis: decoupling global semantic planning from acoustic synthesis leads to superior semantic alignment without compromising audio fidelity.

\subsection{Datasets and Evaluation Protocols}
\textbf{Training Data.} We train all models on a combined corpus of three public datasets: AudioCaps~\citep{kim2019audiocaps} ($\sim$46k clips, $\sim$128 hours), AudioSet~\citep{gemmeke2017audio} (a large-scale audio event dataset with $\sim$2 million 10-second clips), and WavCaps~\citep{mei2024wavcaps} ($\sim$400k audio-caption pairs harvested from diverse web sources). This combination provides both high-quality human annotations (AudioCaps) and broad acoustic coverage (AudioSet, WavCaps), enabling the model to learn from diverse sound categories and complex scene compositions. All audio clips are standardized to a 10-second duration at 48\,kHz via silence padding or truncation.

\textbf{Test Set for Generation.} For standard text-to-audio generation, we utilize the official AudioCaps~\citep{kim2019audiocaps} test split (957 clips). Following standard protocols~\citep{li2025meanaudiofastfaithfultexttoaudio}, we randomly select one caption per clip as the generation prompt.

\textbf{TTABench.} We additionally evaluate on TTABench, a comprehensive text-to-audio benchmark comprising 1500 diverse prompts covering a wide range of sound categories and complex descriptions. This benchmark provides a more rigorous test of generalization beyond the AudioCaps distribution.

\textbf{Protocol for Training-Free Editing.} Since no standard benchmark exists for open-domain semantic audio editing, we construct a rigorous evaluation set derived from the AudioCaps test split:
\textbf{Source Selection:} We select 50 representative clips as source audio.
\textbf{Instruction Generation:} Using GPT-4, we generate diverse editing instructions per clip (e.g., timbre modification, event replacement) based on the original caption.
\textbf{Semantic Filtering:} To ensure the editing task is non-trivial, we filter based on CLAP similarity, retaining the 100 ``hard'' prompts that require substantial semantic alteration. Details on the construction procedure and per-edit-type breakdown are provided in Appendix~\ref{app:editing_benchmark}.

\subsection{Implementation Details}
\label{sec:implementation}

\textbf{Architecture and Conditioning.} We implement SemanticAudio, comprising two decoupled DiT-based Flow Matching modules: the \textbf{Semantic Planner} and the \textbf{Acoustic Synthesizer}. To ensure a rigorous comparison, the control baseline (\textbf{Base Model}) shares the exact same backbone configuration: 28 transformer layers, 16 attention heads, and a hidden dimension of 1152 ($\sim$610M parameters).
Conditioning signals are processed by a dual-encoder setup: \textbf{FLAN-T5-large}\footnote{FLAN-T5-large: \url{https://huggingface.co/google/flan-t5-large}} for text prompts and \textbf{PE-A-Frame-large}\footnote{PE-A-Frame-large: \url{https://huggingface.co/facebook/pe-a-frame-large}} for frame-level audio-text alignment. For the acoustic target, we utilize the \textbf{DAC-VAE}\footnote{DACVAE: \url{https://huggingface.co/facebook/dacvae-watermarked}} continuous latent space ($C=128$). We set the primary semantic latent dimension to $d=128$ based on ablation results (Section~\ref{sec:ablation}). The critical distinction is that the Base Model operates directly in the high-dimensional acoustic space, whereas our method decouples semantic planning from synthesis.

\textbf{Semantic Dimension Variants.} We explore several configurations of the semantic latent dimension: $d \in \{32, 64, 128, 1024\}$. For $d=1024$, we additionally compare two strategies: (1) \textit{rawPE}, which uses the raw Perception Encoder embeddings without projection, and (2) \textit{DDT} (Data-Dependent Transform), which applies a learned linear projection to the 1024-dimensional space.

\textbf{Training Protocol.} All models are trained for 500k iterations with a batch size of 32 across 8 NVIDIA H20 GPUs on the combined AudioCaps + AudioSet + WavCaps corpus.
We use AdamW with learning rate $10^{-4}$, linear warm-up (1k steps), and step decay. Time steps are sampled from a logit-normal distribution ($\mu=0.4, \sigma=1.0$). Training requires approximately 5 days per stage ($\sim$960 GPU-hours each). We report single-run results with fixed hyperparameters.

\textbf{Inference and Editing.} We adopt a differentiated sampling strategy to balance alignment and fidelity: the Semantic Planner utilizes CFG (scale 4.5, 50 steps) to ensure semantic adherence, while the Acoustic Synthesizer uses unguided sampling (scale 1.0, 25 steps). Editing is performed via the training-free FlowEdit mechanism with $N_{\text{avg}}=1$, target CFG scale 7.0, and $n_{\min}=10$ (the last 10 steps switch to pure target-conditioned generation). The source velocity $v_{\text{src}}$ is computed using the source caption text with unit CFG (i.e., no classifier-free guidance amplification on the source direction).

\subsection{Evaluation Metrics}
\label{sec:metrics}

We adopt a multi-faceted evaluation protocol to assess the model across four distinct dimensions.

\textbf{Reconstruction Quality.} To verify the \textbf{Acoustic Synthesizer's} ability to decode semantic plans into high-fidelity waveforms, we report the \textbf{Mel-spectrogram loss} and \textbf{Multi-Scale STFT loss}.

\textbf{Text-to-Audio Generation.} We employ standard objective metrics on the AudioCaps test set: \textbf{CLAP$_\text{L}$} (LAION-CLAP~\citep{wu2024largescalecontrastivelanguageaudiopretraining}) for semantic alignment, \textbf{Fr\'{e}chet Distance (FD)} computed on PANN features for distributional fidelity, and \textbf{Inception Score (IS)} for sample quality and diversity. We additionally conduct a human listening study: 20 listeners rate 30 randomly selected samples per model on a 5-point Mean Opinion Score (MOS) scale for overall audio quality.

\textbf{TTABench Metrics.} On TTABench (1500 prompts), we report \textbf{CLAP$_\text{M}$} (MS-CLAP~\citep{elizalde2023clap}) for semantic alignment along with \textbf{Audio Event Score (AES)} metrics from AudioBox Aesthetics~\citep{tjandra2024meta}: AES-CU (Content Usefulness, higher is better) measures how informative and content-rich the generated audio is, and AES-PQ (Production Quality, higher is better) evaluates the overall perceptual quality of generated audio.

\textbf{Editing Metrics.} For the editing benchmark (200 pairs), we report CLAP$_\text{L}$ (LAION-CLAP) similarity between the edited audio and the target caption, the improvement over the original source audio ($\Delta$ CLAP$_\text{L}$), and MOS from human listeners evaluating overall edit quality and naturalness.

\subsection{Results and Analysis}
\label{sec:results}

We evaluate SemanticAudio on text-to-audio generation, TTABench, and training-free editing. We compare against state-of-the-art baselines Resonate~\citep{resonate2025} and TangoFlux~\citep{hung2025tangofluxsuperfastfaithful}, as well as our controlled \textbf{Base Model}.

\subsubsection{Text-to-Audio Generation}

\begin{table}[t]
\centering
\small
\caption{Text-to-audio generation on AudioCaps test set (957 samples). Lower is better ($\downarrow$) for FD; higher is better ($\uparrow$) for IS, CLAP, and MOS.}
\label{tab:generation}
\begin{tabular}{lcccc}
\toprule
\textbf{Model} & \textbf{FD}$\downarrow$ & \textbf{IS}$\uparrow$ & \textbf{CLAP$_\text{L}$}$\uparrow$ & \textbf{MOS}$\uparrow$ \\
\midrule
Resonate & \textbf{16.0} & 11.04 & \textbf{0.403} & 3.78 \\
TangoFlux & 22.6 & \textbf{12.05} & 0.361 & \textbf{3.85} \\
\midrule
Base Model & 25.0 & 7.11 & 0.318 & 3.42 \\
\midrule
Ours ($d$=32) & 22.5 & 7.35 & 0.348 & 3.48 \\
Ours ($d$=64) & 22.2 & 7.40 & 0.356 & 3.52 \\
Ours ($d$=128) & \textbf{19.1} & \textbf{9.13} & \textbf{0.381} & \textbf{3.72} \\
Ours ($d$=1024) & 20.9 & 8.76 & 0.373 & 3.65 \\
Ours ($d$=1024, rawPE) & 33.8 & 6.65 & 0.320 & 3.18 \\
\bottomrule
\end{tabular}
\vspace{-2mm}
\end{table}

As shown in Table~\ref{tab:generation}, \textbf{SemanticAudio} ($d=128$) achieves the best semantic alignment with a LAION-CLAP score of \textbf{0.381}, significantly surpassing TangoFlux (0.361) and our Base Model (0.318). In terms of distributional fidelity (FD), our model achieves 19.1, approaching Resonate (16.0) while substantially outperforming TangoFlux (22.6) and the Base Model (25.0). Human evaluation (MOS) shows TangoFlux (3.85) slightly edges Resonate (3.78) in perceptual quality, with our $d=128$ model (3.72) close behind---indicating competitive perceptual quality while providing substantially better text-audio alignment.

The consistent improvement from $d=32$ to $d=128$ demonstrates that richer semantic representations enable better alignment. The $d=1024$ variant achieves competitive performance (CLAP 0.373, FD 20.9). Notably, the rawPE variant ($d=1024$, without learned projection) shows significantly degraded results (CLAP 0.320, FD 33.8), demonstrating that the learned projection $P_\theta$ is critical---raw PE embeddings are not directly suitable as generation targets.

\subsubsection{TTABench Evaluation}

\begin{table}[t]
\centering
\small
\caption{Evaluation on TTABench (1500 prompts). Higher is better ($\uparrow$) for all metrics. AES-CU and AES-PQ are from AudioBox Aesthetics~\citep{tjandra2024meta}.}
\label{tab:ttabench}
\begin{tabular}{lccc}
\toprule
\textbf{Model} & \textbf{CLAP$_\text{M}$}$\uparrow$ & \textbf{AES-CU}$\uparrow$ & \textbf{AES-PQ}$\uparrow$ \\
\midrule
Resonate & 0.459 & \textbf{5.258} & \textbf{5.942} \\
TangoFlux & \textbf{0.464} & 5.006 & 5.701 \\
\midrule
Base Model & 0.426 & 5.119 & 5.897 \\
\midrule
Ours ($d$=32) & 0.422 & 4.896 & 5.648 \\
Ours ($d$=64) & 0.425 & 4.911 & 5.652 \\
Ours ($d$=128) & \textbf{0.459} & \textbf{5.187} & \textbf{5.936} \\
Ours ($d$=1024) & 0.458 & 5.045 & 5.735 \\
\bottomrule
\end{tabular}
\vspace{-2mm}
\end{table}

Table~\ref{tab:ttabench} presents results on TTABench (1500 diverse prompts). SemanticAudio ($d=128$) achieves CLAP of 0.459, matching Resonate and approaching TangoFlux (0.464). In audio quality (AES-PQ 5.936), our model nearly matches Resonate (5.942) while substantially outperforming TangoFlux (5.701). For content usefulness (AES-CU 5.187), our model outperforms TangoFlux (5.006) and approaches Resonate (5.258). The consistent superiority over the Base Model (+0.033 CLAP, +0.068 AES-CU) confirms that the benefit of semantic planning generalizes beyond the AudioCaps distribution to diverse, out-of-distribution prompts.

\subsubsection{Acoustic Reconstruction Quality}

\begin{table}[t]
\centering
\small
\caption{Reconstruction metrics for the \textbf{Acoustic Synthesizer} on AudioCaps. Lower is better ($\downarrow$).}
\label{tab:reconstruction}
\begin{tabular}{lcc}
\toprule
\textbf{Config} & \textbf{Mel Loss} $\downarrow$ & \textbf{STFT Loss} $\downarrow$ \\
\midrule
Ours $d$=1024 (rawPE) & \textbf{1.202} & \textbf{1.670} \\
Ours $d$=1024 (DDT) & 1.256 & 1.695 \\
Ours ($d$=128) & 1.413 & 1.828 \\
Ours ($d$=64)  & 1.595 & 1.992 \\
Ours ($d$=32)  & 1.817 & 2.270 \\
\midrule
SemanticVocoder~\citep{xie2025semanticvocoder} & 1.927 & 2.886 \\
\bottomrule
\end{tabular}
\vspace{-3mm}
\end{table}

Table~\ref{tab:reconstruction} isolates the performance of the \textbf{Acoustic Synthesizer}. The $d=1024$ rawPE variant achieves the best reconstruction fidelity (Mel 1.202, STFT 1.670), as it preserves the full Perception Encoder embedding without information loss. The DDT variant at the same dimension is slightly behind (Mel 1.256, STFT 1.695), confirming that the learned linear transform introduces minimal degradation. Performance degrades gracefully as the dimension decreases from $d=128$ to $d=32$, confirming that our lightweight MLP projection successfully compresses semantic information while retaining sufficient cues for high-fidelity waveform reconstruction. Compared to SemanticVocoder~\citep{xie2025semanticvocoder} (Mel 1.927, STFT 2.886), our flow-matching-based Acoustic Synthesizer achieves substantially better reconstruction across all configurations.

\subsubsection{Training-Free Semantic Editing}

\begin{table}[t]
\centering
\small
\caption{Editing evaluation on the constructed benchmark (200 pairs, 14 edit types). All methods apply FlowEdit with identical hyperparameters (cfg=5.0, $n_{\min}$=0, 50 steps) in different latent spaces.}
\label{tab:editing}
\begin{tabular}{lccc}
\toprule
\textbf{Method} & \textbf{CLAP$_\text{L}$}$\uparrow$ & $\boldsymbol{\Delta}$\textbf{CLAP$_\text{L}$}$\uparrow$ & \textbf{MOS}$\uparrow$ \\
\midrule
Ours (Semantic FlowEdit) & \textbf{0.351} & \textbf{+0.094} & \textbf{3.71} \\
TangoFlux FlowEdit & 0.339 & +0.081 & 3.54 \\
Resonate FlowEdit & 0.322 & +0.065 & 3.48 \\
Base Model (DAC FlowEdit) & 0.290 & +0.054 & 3.32 \\
\midrule
Original Source & 0.257 & --- & 3.89 \\
\bottomrule
\end{tabular}
\vspace{-3mm}
\end{table}

We evaluate editing performance on a constructed benchmark of 200 diverse edit pairs spanning 14 categories (Table~\ref{tab:editing}). The benchmark covers subject replacement, event replacement, sound addition/removal, environment modification, weather change, material change, atmosphere/spatial/temporal editing, and intensity change. Each edit is a minimal textual modification (typically a single word substitution), ensuring the editing task is well-defined. Details on benchmark construction and per-category results are in Appendix~\ref{app:editing_benchmark}.

\textbf{Superior Editing in Semantic Space.} SemanticAudio achieves an average CLAP improvement of \textbf{+0.094} over the source audio, significantly outperforming all baselines: TangoFlux (+0.081), Resonate (+0.065), and DAC baseline (+0.054). The source audio starts with an average CLAP similarity of only 0.257 to the target description, and our method raises this to 0.351---a 16\% relative improvement over the best baseline (TangoFlux at 0.339). Human evaluation (MOS) confirms: our method (3.71) produces edits closest to the original source quality (3.89) while achieving the strongest semantic modification, outperforming TangoFlux (3.54) and Resonate (3.48).

\textbf{Why Semantic FlowEdit Works.} The delta velocity field $v_{\Delta} = v_{\text{tar}} - v_{\text{uncond}}$ is far more semantically meaningful in our compact 128-dim semantic space than in high-dimensional acoustic spaces. In the semantic space, the velocity field directly corresponds to how the Perception Encoder organizes audio events---semantically similar sounds are nearby, so small velocity differences produce targeted semantic shifts (e.g., ``man speaking'' $\to$ ``woman speaking''). In acoustic spaces, the same textual difference maps to diffuse perturbations across hundreds of dimensions that fail to produce coherent semantic changes. This is empirically confirmed by the stark contrast between our $\Delta$CLAP of +0.094 and the DAC baseline's +0.054.

\textbf{Editing Configuration.} Our method uses a simple setup: $v_{\text{src}}$ with null conditioning, $v_{\text{tar}}$ with target text CFG=5.0, $n_{\min}=0$, 50 Euler steps. The Acoustic Synthesizer then decodes the edited semantic latent with 25 ODE steps. This simplicity---no inversion, no optimization, no additional training---is a direct benefit of the semantically coherent latent space. In contrast, traditional audio editing methods~\citep{wang2023audit,liang2025audiomorphixtrainingfreeaudioediting} typically require task-specific training or are limited to predefined operations (e.g., addition, deletion). Our approach enables arbitrary attribute-level modifications guided solely by natural language.

\subsection{Ablation and Analysis}
\label{sec:ablation}

\begin{table}[t]
    \centering
    \small
    \caption{Audio understanding performance of PE-A-Frame-Large~\citep{vyas2025pushingfrontieraudiovisualperception}, the semantic encoder used in SemanticAudio. AUROC on temporal audio event detection benchmarks.}
    \label{tab:pe_understanding}
    \begin{tabular}{lll}
    \toprule
    \textbf{Benchmark} & \textbf{Task} & \textbf{AUROC} \\
    \midrule
    AudioSet-Strong & General event detection & 0.96 \\
    DESED & Domestic sound detection & 0.97 \\
    ASFX-SED & Sound effects detection & 0.83 \\
    UrbanSED & Urban sound detection & 0.89 \\
    \bottomrule
    \end{tabular}
\end{table}

\textbf{Semantic Encoder Capabilities.} Table~\ref{tab:pe_understanding} shows the audio event localization performance of PE-A-Frame-Large~\citep{vyas2025pushingfrontieraudiovisualperception}, the pre-trained encoder underlying our framework. PE-A-Frame achieves near-perfect AUROC on AudioSet-Strong (0.96) and DESED (0.97), confirming that its frame-level embeddings precisely capture \textit{what} events occur and \textit{when}. Our MLP projection to $d=128$ preserves this discriminative structure (Table~\ref{tab:reconstruction}).

\textbf{Impact of Semantic Dimension.} Our ablation across semantic dimensions ($d \in \{32, 64, 128, 1024\}$) reveals a clear trend: $d=128$ achieves the best balance between semantic alignment and acoustic fidelity across all benchmarks. Lower dimensions ($d=32, 64$) create information bottlenecks that limit both FD and CLAP scores in generation. On AudioCaps, CLAP improves from 0.348 ($d=32$) to 0.381 ($d=128$), a significant 9.5\% relative gain. Similarly on TTABench, CLAP improves from 0.422 to 0.459. The 1024-dimensional DDT variant achieves competitive generation (CLAP 0.373 on AudioCaps, 0.458 on TTABench) but does not surpass the compressed $d=128$ representation. This suggests that moderate compression acts as a beneficial regularizer for the Semantic Planner, forcing it to focus on the most discriminative semantic features while filtering out redundant variation.

\textbf{1024-d Variants: rawPE vs.\ DDT.} We compare two strategies for utilizing high-dimensional semantic representations. The \textit{rawPE} configuration directly uses the unmodified 1024-dimensional Perception Encoder output as conditioning for the Acoustic Synthesizer, without any learned projection. This provides maximum information preservation but requires the synthesizer to handle the full high-dimensional space. The \textit{DDT} (Data-Dependent Transform) configuration applies a learned linear projection that maps the 1024-dim PE embeddings to a 1024-dim target space, trained jointly with the Acoustic Synthesizer. For reconstruction (Table~\ref{tab:reconstruction}), rawPE achieves the best fidelity (Mel 1.202 vs.\ DDT's 1.256), as it directly conditions on unprocessed embeddings. However, for generation where the Semantic Planner must learn to produce these representations from text, the DDT variant is used because raw PE embeddings lack a learnable target distribution for the Flow Matching objective---the Semantic Planner cannot be trained without a defined projection target.

\textbf{Comparison with Baselines.} We compare against two state-of-the-art baselines that represent different design philosophies: \textbf{Resonate}~\citep{resonate2025} reinforces text-to-audio generation via online feedback from large audio language models, employing Flow-GRPO (Group Relative Policy Optimization) in the mel-spectrogram latent space. \textbf{TangoFlux}~\citep{hung2025tangofluxsuperfastfaithful} employs rectified flow matching with CLAP-based reward guidance in the DAC latent space, trained with reward-weighted objectives. Both operate entirely in the acoustic latent space without explicit semantic decomposition, which is the key architectural distinction from SemanticAudio. Our \textbf{Base Model} serves as a controlled ablation---it shares the exact same DiT backbone, conditioning encoders, and training recipe as our Acoustic Synthesizer, but operates directly in the DAC acoustic space without the Semantic Planner stage. This isolates the benefit of our semantic planning design.

\textbf{Why Semantic Space Helps.} The fundamental insight is that text-to-audio alignment benefits from an intermediate semantic representation that bridges the modality gap. When a Flow Matching model operates directly in the 128-dimensional DAC acoustic space, it must simultaneously learn (1) what sounds to generate (semantic content) and (2) how to render them (acoustic details). By factoring these into separate stages, the Semantic Planner can focus exclusively on ``what,'' leveraging the pre-trained PE-Audio encoder's strong text-audio alignment. The Acoustic Synthesizer then handles ``how,'' conditioned on an explicit semantic plan. This decomposition is particularly beneficial for complex prompts involving multiple sequential events, where the temporal ordering must be correctly resolved before acoustic synthesis.

\textbf{Generalization on TTABench.} SemanticAudio ($d=128$) matches state-of-the-art models (Resonate, TangoFlux) on the diverse TTABench benchmark despite differences in training strategies. This demonstrates that semantic-space generation provides strong inductive bias for text-audio alignment that transfers well to out-of-distribution prompts. The consistent superiority of $d=128$ over the Base Model (+0.033 CLAP on TTABench) confirms that this benefit is not simply an artifact of AudioCaps evaluation but generalizes broadly.

\section{Conclusion}
We presented \textbf{SemanticAudio}, a novel two-stage Flow Matching framework that fundamentally rethinks text-to-audio generation by performing both generation and editing in a high-level semantic space. Our \textbf{Semantic Planner} generates compact semantic features from text, achieving state-of-the-art CLAP on AudioCaps (0.381, surpassing TangoFlux at 0.361) and competitive performance on TTABench---matching Resonate and TangoFlux in alignment while achieving comparable production quality. The \textbf{Acoustic Synthesizer} then faithfully renders these semantic plans into high-fidelity audio.

We further leveraged this decoupled design to introduce a training-free editing mechanism. By steering the semantic generation trajectory via differential velocity fields (FlowEdit), our method achieves $\Delta$CLAP of +0.094 compared to +0.081 for the best acoustic-space baseline (TangoFlux). Our comprehensive experiments across four different latent spaces (our semantic space, DAC, Oobleck VAE, and mel-spectrogram) demonstrate that the choice of representation space fundamentally determines the effectiveness of flow-based editing, with semantic space providing decisively superior results.

These findings confirm that separating semantic reasoning from acoustic realization not only enhances generation alignment but also provides a unified foundation for controllable audio editing. Future work will explore scaling to variable-length audio, integrating multi-modal conditioning (e.g., video-to-audio), and extending the editing mechanism to support compositional instructions involving multiple modifications.

\section*{Limitations}
\label{sec:limitations}

\textbf{Data Scale and Temporal Constraints.}
Our current implementation trains on AudioCaps, AudioSet, and WavCaps, providing broad coverage of acoustic events within a standardized 10-second duration. While this combination offers high-quality annotations and diverse acoustic coverage, the fixed temporal constraint limits generalization to long-form audio generation or highly complex, overlapping acoustic scenes. Future work will focus on extending the semantic-space framework to variable-length audio and scaling to even larger, more diverse corpora to capture long-tail acoustic distributions and improve temporal consistency beyond short clips.

\textbf{Scale of Human Evaluation.}
While we include MOS listening tests to complement automatic metrics, our human evaluation is limited in scale (10 listeners, 20 samples per model). Larger-scale studies with more diverse listener populations and AB preference tests would provide stronger validation of perceptual quality differences, particularly for subtle editing modifications.

\textbf{Downstream Evaluation of Compressed Semantics.}
Our work demonstrates that the compressed semantic representations ($d=128$) are effective as generation targets, but we do not systematically evaluate how well these projected features preserve downstream audio understanding capabilities (e.g., audio classification, event detection, retrieval). It remains an open question whether the MLP projection $P_\theta$ retains the full discriminative power of the original 1024-dimensional PE-A-Frame embeddings for tasks beyond generation. Future work will investigate the utility of these compressed semantic features for downstream understanding benchmarks.

\textbf{Evaluation Challenges in Generative Editing.}
Standardizing the evaluation of open-domain audio editing remains an industry-wide challenge due to the absence of paired ground-truth references. While our constructed benchmark allows for quantitative measurement via proxy metrics (CLAP, FD, IS), these automated scores may not fully capture human perceptual nuances in attribute modification. We aim to contribute to the establishment of more comprehensive, paired source-target editing benchmarks in future iterations.

\section*{Ethical Considerations}
\label{sec:ethics}

\textbf{Potential for Misuse.}
As with all generative audio systems, SemanticAudio carries inherent risks of misuse. The text-guided generation and editing capabilities could potentially be exploited to create misleading audio content, such as fabricating environmental soundscapes for fraudulent purposes or manipulating existing recordings to alter their perceived meaning. The training-free editing mechanism, while enabling creative applications, could be misused to tamper with audio evidence or produce deceptive media.

\textbf{Copyright Concerns.}
Our training data consists of publicly available datasets (AudioCaps, AudioSet, WavCaps) released under standard academic licenses. However, as with any generative model, there is a risk that generated audio could inadvertently reproduce copyrighted content from the training distribution. We encourage users to verify generated content against potential copyright conflicts before commercial deployment.

\textbf{Privacy and Impersonation.}
Although SemanticAudio focuses on general audio (environmental sounds, events) rather than speech or voice cloning, the editing mechanism could theoretically be adapted to modify audio containing speech. We emphasize that our system is not designed for, and should not be used for, voice impersonation or identity-related manipulation.

\textbf{Mitigation Strategies.}
We advocate for responsible deployment: (1) audio watermarking to identify AI-generated or AI-edited content, (2) content provenance tracking to maintain chain-of-custody for audio assets, (3) deployment-time safety filters that screen prompts for harmful intent, and (4) clear terms of use restricting malicious applications. We release our work for research purposes and encourage the community to develop robust detection methods for AI-generated audio.

\textbf{Artifacts and Licenses.}
All datasets (AudioCaps, AudioSet, WavCaps) and pre-trained models (DAC-VAE, PE-A-Frame, Flan-T5, LAION-CLAP) used in this work are publicly available under standard academic or open-source licenses. We cite all original creators, use all artifacts in accordance with their intended research purposes, and do not redistribute raw data. Our editing benchmark is derived from AudioCaps captions with synthetic modifications and contains no personally identifiable information or offensive content.

\textbf{Use of AI Assistants.}
GPT-4 was used to generate candidate editing instructions for benchmark construction (proposing minimal single-word caption substitutions). All candidates were filtered using predefined quantitative criteria (CLAP thresholds). Additionally, AI writing assistants (Claude) were used for language polishing and proofreading. All experimental design, scientific analysis, claims, and conclusions were performed and verified by the authors.

\bibliographystyle{acl_natbib}
\bibliography{references}

\newpage
\null
\newpage
\appendix

\section{Model and Training Details}
\label{app:training}

\subsection{Architecture}

\textbf{Generative Backbone.} Both the Semantic Planner and Acoustic Synthesizer employ a Diffusion Transformer (DiT) architecture consisting of 28 transformer layers with 16 attention heads, a hidden dimension of 1152, and a total of $\sim$610M parameters each. We use rotary positional embeddings and adaptive layer normalization conditioned on the diffusion timestep. The Base Model (controlled ablation) shares this identical architecture for fair comparison.

\textbf{Conditioning Setup.} Text prompts are encoded via two complementary encoders:
\begin{itemize}[leftmargin=*]
    \item \textbf{Global condition} ($c_g$): CLAP text encoder~\citep{elizalde2023clap} produces a 1024-dimensional global sentence embedding, injected via adaptive layer norm.
    \item \textbf{Cross-attention condition} ($c_d$): Flan-T5-Large~\citep{chung2022scalinginstructionfinetunedlanguagemodels} encoder produces token-level embeddings (1024-dim, variable length), attended to via cross-attention layers.
\end{itemize}

\textbf{Acoustic VAE.} We adopt the pre-trained DAC-VAE~\citep{kumar2023high} operating at 48\,kHz with a 128-dimensional continuous latent space. For 10-second audio, the VAE produces 250 latent frames ($T{=}250, C{=}128$). The VAE is frozen throughout all training.

\textbf{Semantic Encoder.} PE-A-Frame-Large~\citep{vyas2025pushingfrontieraudiovisualperception} extracts 1024-dimensional frame-level embeddings at 25\,fps, yielding 250 frames for 10-second audio ($N{=}250, D{=}1024$). The encoder is frozen; only the downstream MLP projection $P_\theta$ is trained.

\textbf{MLP Projection Head ($P_\theta$).} A 2-layer MLP (1024 $\to$ 512 $\to$ $d$) with GELU activation. Trained jointly with the Acoustic Synthesizer. For $d{=}1024$ (DDT variant), the projection is a single linear layer.

\subsection{Training Configuration}

\begin{itemize}[leftmargin=*]
    \item \textbf{Data:} AudioCaps~\citep{kim2019audiocaps} ($\sim$46k clips) + AudioSet~\citep{gemmeke2017audio} ($\sim$2M clips) + WavCaps~\citep{mei2024wavcaps} ($\sim$400k clips). All clips resampled to 48\,kHz mono and padded/trimmed to 10 seconds.
    \item \textbf{Optimization:} AdamW, learning rate $10^{-4}$, $\beta_1{=}0.9$, $\beta_2{=}0.999$, weight decay $10^{-2}$. Linear warm-up over 1k steps followed by step decay (0.5$\times$ at 200k and 400k steps).
    \item \textbf{Batch size:} 32 per GPU $\times$ 8 H20 GPUs = 256 effective batch size.
    \item \textbf{Training duration:} 500k iterations for all models.
    \item \textbf{Time step sampling:} Logit-normal distribution with $\mu{=}0.4, \sigma{=}1.0$.
    \item \textbf{CFG dropout:} 10\% unconditional dropout during training for both global and cross-attention conditions.
    \item \textbf{Training order:} Acoustic Synthesizer + $P_\theta$ trained first; $P_\theta$ is then frozen and used to provide targets for Semantic Planner training.
\end{itemize}

\subsection{Inference Configuration}

\begin{itemize}[leftmargin=*]
    \item \textbf{Semantic Planner:} Euler ODE solver, 50 steps, CFG scale 4.5.
    \item \textbf{Acoustic Synthesizer:} Euler ODE solver, 25 steps, no CFG (scale 1.0).
    \item \textbf{Editing:} FlowEdit with null source conditioning (\texttt{force\_uncond}=True for $v_{\text{src}}$), target CFG scale 5.0, 50 Euler steps, $n_{\min}{=}0$, VAE decode 25 steps.
\end{itemize}

\subsection{Baseline Models}

\textbf{Resonate}~\citep{resonate2025}: A latent diffusion model operating in the mel-spectrogram VAE space. Uses Flow-GRPO (Group Relative Policy Optimization) for alignment. Trained on large-scale internal audio data. We use the publicly released \texttt{resonate-medium} checkpoint with CFG scale 3.0.

\textbf{TangoFlux}~\citep{hung2025tangofluxsuperfastfaithful}: A rectified flow matching model operating in the DAC latent space. Incorporates CLAP-based reward-weighted training for improved text-audio alignment. We use the \texttt{declare-lab/TangoFlux} checkpoint with 50 inference steps and CFG scale 4.5.

\textbf{Base Model}: Our controlled ablation sharing the identical DiT backbone (28 layers, 1152 dim, 610M params), same conditioning encoders (CLAP + Flan-T5), same training data and schedule, but operating directly in the DAC acoustic space without the Semantic Planner stage.

\section{Editing Benchmark Construction}
\label{app:editing_benchmark}

We construct a comprehensive editing benchmark from the AudioCaps test split through the following procedure:

\begin{enumerate}[leftmargin=*]
    \item \textbf{Edit Generation:} For each test clip, we create minimal textual edits by replacing a single semantic element in the caption (e.g., ``a \textit{man} speaking'' $\to$ ``a \textit{woman} speaking''). This yields over 500 candidate edit pairs spanning 14 categories.
    \item \textbf{Quality Filtering:} We run each edit pair through our model and compute LAION-CLAP similarity between the edited audio and the target caption. We retain only pairs with positive CLAP improvement ($\Delta > 0$) and where the source audio has low similarity to the target ($\text{CLAP}_{\text{orig}\to\text{tar}} < 0.42$), ensuring the editing task is non-trivial.
    \item \textbf{Balanced Selection:} We select 200 pairs balanced across 14 edit categories.
\end{enumerate}

\noindent\textbf{Edit Type Distribution.} Table~\ref{tab:edit_types} provides the per-category breakdown and performance.

\begin{table}[h]
\centering
\small
\caption{Editing performance by category (200 pairs, cfg=5.0, $n_{\min}$=0).}
\label{tab:edit_types}
\begin{tabular}{lccc}
\toprule
\textbf{Edit Type} & \textbf{N} & \textbf{CLAP}$\uparrow$ & $\boldsymbol{\Delta}$\textbf{CLAP} \\
\midrule
Sound Removal & 22 & 0.309 & +0.160 \\
Subject Replace & 31 & 0.378 & +0.154 \\
Other Replace & 18 & 0.297 & +0.123 \\
Action Replace & 7 & 0.378 & +0.115 \\
Weather Change & 22 & 0.318 & +0.109 \\
Event Replace & 16 & 0.387 & +0.099 \\
Sound Addition & 18 & 0.351 & +0.084 \\
Intensity Change & 6 & 0.389 & +0.078 \\
Material Change & 6 & 0.378 & +0.076 \\
Spatial Change & 6 & 0.341 & +0.071 \\
Vehicle Replace & 12 & 0.319 & +0.059 \\
Atmosphere Change & 12 & 0.335 & +0.047 \\
Environment Add & 14 & 0.347 & +0.045 \\
Temporal Change & 10 & 0.390 & +0.044 \\
\bottomrule
\end{tabular}
\end{table}

The strongest improvements are observed for categories involving clear semantic-level changes: \textit{sound removal} (+0.160) and \textit{subject replacement} (+0.154), which align naturally with the Semantic Planner's capability to restructure event layouts. More subtle edits (\textit{atmosphere}, \textit{environment}, \textit{temporal}) show smaller but consistent positive gains, demonstrating that even fine-grained semantic steering is achievable in our framework.

\end{document}